\documentclass[]{aa}  

\usepackage{graphicx}
\usepackage{txfonts}
\usepackage{amsmath}

\usepackage[colorlinks=true,linkcolor=magenta,citecolor=blue,filecolor=black,urlcolor=cyan,final=true]{hyperref}
\def\code#1{\texttt{#1}}
\newcommand{\unit}[1]{\ensuremath{\, \mathrm{#1}}}

\begin{document}

   \title{High-Energy Insights from an Escaping Coronal Mass Ejection with Solar Orbiter/STIX Observations}

%   \subtitle{I. Overviewing the $\kappa$-mechanism}

   \author{L. A. Hayes
          \inst{1}
          \and
          S\"am Krucker\inst{2, 3}
          \and 
          H. Collier\inst{2, 4}
          \and
          D. Ryan\inst{2}
          }

   \institute{
              %\email{laura.hayes@esa.int}
         %\and
             European Space Agency (ESA), European Space Research and Technology Centre (ESTEC), Keplerlaan 1, 2201 AZ Noordwijk, The Netherlands \email{laura.hayes@esa.int, lauraannhayes@gmail.com}
         \and 
             University of Applied Sciences and Arts Northwestern Switzerland, Bahnhofstrasse 6, 5210 Windisch, Switzerland
         \and
             Space Sciences Laboratory, University of California, 7 Gauss Way, 94720 Berkeley, USA 
        \and 
            ETH Zürich, Rämistrasse 101, 8092 Zürich Switzerland}
   \date{Received xxx; accepted xxx}

% \abstract{}{}{}{}{} 
% 5 {} token are mandatory
 
  \abstract
   {Solar eruptive events, including solar flares and coronal mass ejections (CMEs), are typically characterised by energetically significant X-ray emissions from flare-accelerated electrons and hot thermal plasmas. However, the intense brightness of solar flares often overshadows high-coronal X-ray emissions from the associated eruptions due to the limited dynamic range of current instrumentation. Occulted events, where the main flare is blocked by the solar limb, provide an opportunity to observe and analyse the X-ray emissions specifically associated with CMEs.}
   {This study investigates the X-ray and extreme ultraviolet (EUV) emissions associated with a large filament eruption and CME that occurred on February 15, 2022. This event was highly occulted from the three vantage points of Solar Orbiter ($\sim$45$^{\circ}$ behind the limb), Solar–TErrestrial RElations Observatory (STEREO-A), and Earth.}
   { We utilised X-ray observations from the Spectrometer/Telescope for Imaging X-rays (STIX) and EUV observations from the Full Sun Imager (FSI) of the Extreme Ultraviolet Imager (EUI) on-board Solar Orbiter, supplemented by multi-viewpoint observations from STEREO-A/Extreme-UltraViolet Imager (EUVI). This enabled a comprehensive analysis of the X-ray emissions in relation to the filament structure observed in EUV. We used STIX's imaging and spectroscopy capabilities to characterise the X-ray source associated with the eruption.}
   {Our analysis reveals that the X-ray emissions associated with the occulted eruption originated from an altitude exceeding 0.3~\(R_\odot\) above the main flare site. The X-ray time-profile showed a sharp increase and exponential decay, and consisted of both a hot thermal component at 17$\pm$2~MK and non-thermal emissions ($>11.4\pm0.2$~keV) characterised by an electron spectral index of 3.9$\pm$0.2. Imaging analysis showed an extended X-ray source that coincided with the EUV emission as observed from EUI, and was imaged until the source grew to a size larger than the imaging limit of STIX (180\arcsec).}
   {Filament eruptions and associated CMEs have hot and non-thermal components and the X-ray emissions are energetically significant. The findings demonstrate that STIX combined with EUI provides a unique and powerful tool for examining the energetic properties of the CME component of solar energetic eruptions. Multi-viewpoint and multi-instrument observations are crucial for revealing such energetically significant sources in solar eruptions that might otherwise remain obscured.
}


   \maketitle
%
%-------------------------------------------------------------------

\section{Introduction}
Solar eruptive events, which include solar flares, coronal mass ejections (CMEs), and occasionally associated solar energetic particle (SEP) events, are fundamental drivers of space weather and constitute the most energetically significant processes within our solar system. These phenomena are driven by the rapid release of free magnetic energy in the Sun's corona. According to the standard model of solar eruptive events, the energy release is predominantly driven by large-scale magnetic reconnection beneath an erupting magnetic flux rope or filament, leading to the flare and CME.  A fundamental challenge in solar physics is understanding how this released magnetic energy is partitioned and transformed into other forms, such as heat and particle acceleration. 

X-ray observations of solar eruptive events provide one of the most direct diagnostics of the hottest flare plasma and the non-thermal flare-accelerated electrons. Typically in solar eruptive events, these observations reveal bright hard X-ray footpoints within the chromosphere, as well as hot flare loops rooted in these footpoints that often form an arcade in the lower corona \citep{fletcher2011, benz2017}. Although it is theorised - and occasionally observed - that hard X-ray emissions also originate from higher in the corona near the acceleration regions or in conjunction with CME eruptions, detecting these emissions presents a significant challenge. Since the intensity of bremsstrahlung X-ray emission depends on the ambient density, emission from the denser lower solar atmosphere dominates and, given the limited dynamic range of current and past X-ray imaging instruments, usually obscures higher altitude sources. Nevertheless, in certain exceptional circumstances \citep[e.g.][]{masuda1994, krucker_battaglia}, or during times when the solar flare footpoints are occulted behind the solar limb, coronal hard X-ray emissions become detectable, offering rare glimpses into high-corona processes \citep[e.g.][]{krucker_lin2008, effenberger, lastufka2019} .

In only a handful of cases where the main flare is either fully or partially occulted by the solar limb, high-coronal hard X-ray emissions associated with CME eruptions have been reported, with both Yohkoh and RHESSI \citep{hudson2001, krucker_2007, glesener2013, lastufka2019}. In these reported cases, both a non-thermal and hot thermal component of the erupting flux ropes were reported, suggesting that energetic electrons and hot plasma are present in the CME erupting filament. More recently with observations from the Expanded Owens Valley Solar Array (EOVSA), \cite{chen2020} reported for the first time non-thermal microwave emission in the anchor points of the erupting filament, further suggesting the presence of accelerated electrons trapped within erupting flux ropes. With recent X-ray observations from the Spectrometer Telescope for Imaging X-rays (STIX) \citep{krucker2020} on-board Solar Orbiter \citep{mueller_2020}, \cite{stiefel_2023}, reported the first observations on non-thermal emissions in the flux-rope filament footpoints (albeit in the chromosphere) - where flare-accelerated energetic electrons trapped in the flux-rope presumably precipitated to the denser anchor points of the flux-rope to produce X-ray emission. Understanding the nature and energetics of non-thermal emission from erupting filaments is an important aspect of understanding particle acceleration and energy release in solar eruptive events.

In this paper, we present hard X-ray observations of a filament eruption that occurred on February 15, 2022, captured by the Spectrometer/Telescope for Imaging X-rays (STIX) and the Full Sun Imager (FSI) of the Extreme Ultraviolet Imager (EUI) on-board Solar Orbiter \citep{rochus2020}. STIX offers hard X-ray imaging and spectroscopy, measuring X-rays in the 4-150 keV energy range. Our analysis includes both the spatial and spectral aspects of the X-ray observations, revealing that the eruption exhibited both hot and non-thermal components. This work highlights the unique observational opportunities provided by STIX and other instruments on-board Solar Orbiter. Section 2 presents an overview of the event, as well as more detailed X-ray imaging and spectroscopic analyses, while Section 3 discusses the implications of this work and provides our conclusions. 

%--------------------------------------------------------------------
\section{Observations}
\subsection{Event overview}
\begin{figure}
    \centering
    \includegraphics[width=0.38\textwidth]{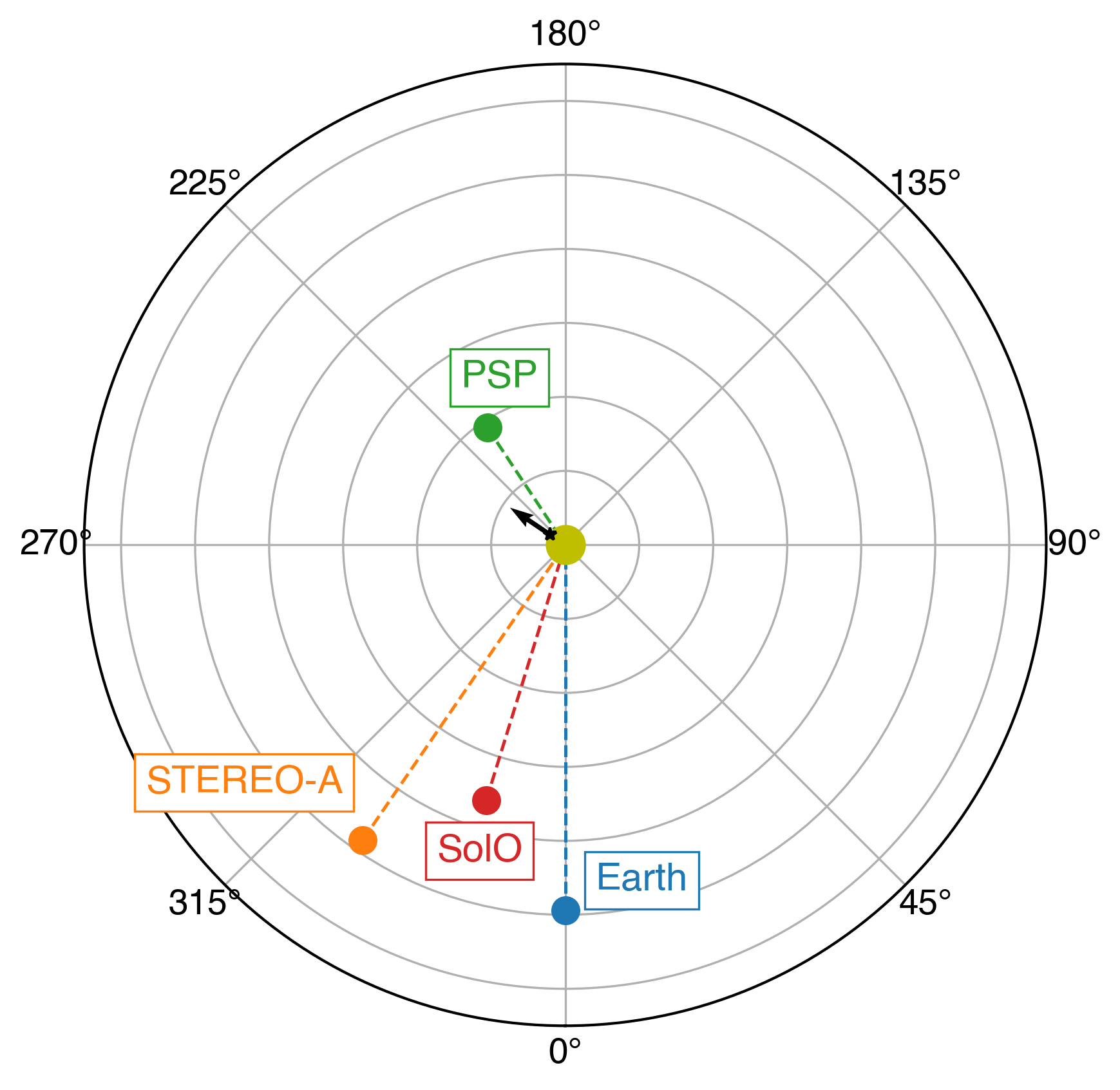}
    \caption{A plot of the relative positions of the Solar Orbiter, Earth, STEREO-A, and PSP at the time of the CME eruption. The black arrow marks the estimated longitude of the eruption.}
    \label{fig:positions}
\end{figure}

On February 15, 2022, a large filament eruption took place off the eastern limb of the Sun, captured from three distinct observational vantage points: Earth, Solar–TErrestrial RElations Observatory (STEREO-A), and Solar Orbiter. The main body of the event was obscured by the solar limb from all three locations, providing a unique perspective on the eruption. Figure~\ref{fig:positions} illustrates the positions of the spacecraft relative to the Sun, and the black arrow in the figure indicates the filament eruption's location and trajectory, determined to be $\sim$59$^{\circ}$ behind the limb as viewed from Earth. The eruption was clearly captured by the large field of view of the EUI/FSI telescope on-board Solar Orbiter, where parts of the filament eruption were tracked out to 6~\(R_\odot\) in 304~\AA\ \citep[see][]{mierla_2022}. The eruption was also associated with a fast CME ($\sim$2200~km/s) and an interplanetary CME-driven shock which caused a widespread solar energetic particle event that was detected by in-situ instrument on several spacecraft throughout the heliosphere \citep[see][]{palmerio2024, giacalone_2023, khoo2024}. Such fast CME's have been known to produce hard X-ray signatures in the high corona \citep{krucker_2007}.

\begin{figure}
    \centering
    \includegraphics[width=0.45\textwidth]{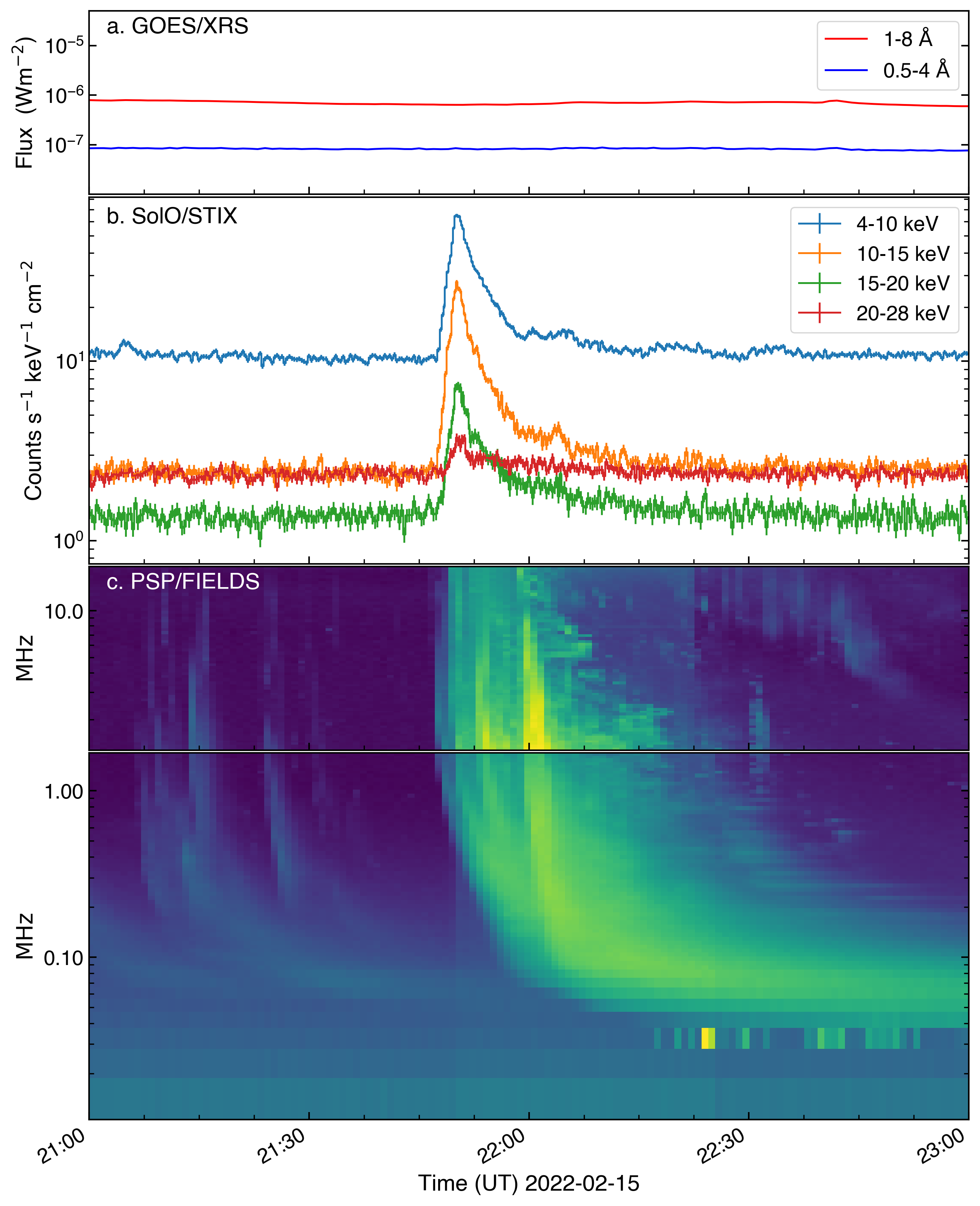}
    \caption{The X-ray and radio observations of the eruption as observed by Earth, Solar Orbiter and PSP. The GOES XRS 1--8~\AA\ and 0.5--4~\AA\ channels are shown in panel a. The STIX lightcurves in 4 energy bands (flux summed over these energies) is shown in panel (b), where clearly X-ray emission associated with the eruption can be observed. In panel (c), the dynamic spectra from HFR and LFR from PSP/FIELDS is plotted, where the Type III radio bursts associated with the event is observed. Langmuir waves are also observed at $\sim$0.05MHz after around 22:20 UT. The times here are all adjusted to UT at PSP location.}
    \label{fig:overview}
\end{figure}

Figure~\ref{fig:overview} presents X-ray and radio measurements taken from various viewpoints during the filament eruption. Panel a demonstrates that, from Earth's perspective, there was no detectable enhancement in X-ray emission across the channels of the GOES-16 X-ray Sensor (XRS). Conversely, panel b shows that from the perspective of Solar Orbiter, STIX captured clear X-ray emissions reaching up to 28~keV, even though the event was also occulted behind the limb from this viewpoint. Panel c features radio observations from the Parker Solar Probe (PSP)/FIELDS \citep{bale_2016} across both the Low Frequency Receiver (LFR; bandwidth 10.5 kHz–1.7~MHz) and the High Frequency Receiver (HFR; 1.3~MHz–19.2~MHz). During the eruption, pronounced radio bursts characteristic of Type-III signatures were detected, their onset aligning with the peak of the X-ray emissions captured by STIX. These bursts indicate the presence of impulsively accelerated electron beams travelling from the solar atmosphere's lower layers at the flare site into interplanetary space along open magnetic field lines. Additionally, Langmuir wave signatures were identified in the dynamic spectrum at frequencies around 0.05~MHz after approximately 22:30 UT, indicating in-situ measurements of these Type-III generating electron beams by PSP and suggesting a magnetic connection between the spacecraft and the event source.

\begin{figure*}
    \centering
    \includegraphics[width=0.9\textwidth]{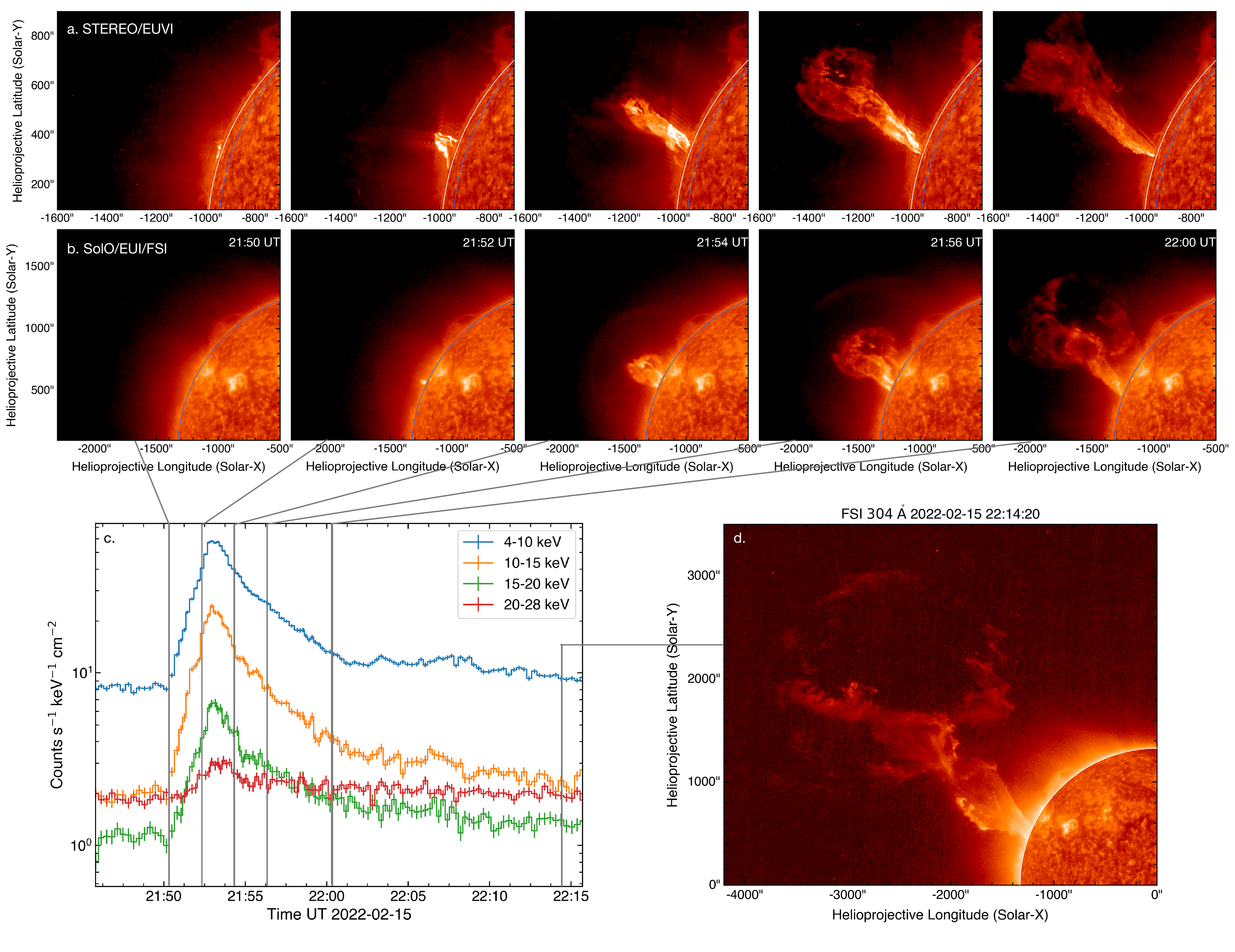}
    \caption{An overview of the EUV observations in 304~\AA\ from the different vantage points of STEREO/EUVI and Solar Orbiter/EUI/FSI. Panel (a) and panel (b) shows five consecutive images from EUVI and FSI, respectively. The images of EUVI and EUI are matched to the closest time, taking into light-travel time, and are given in the time at Solar Orbiter. They are not reprojected to one viewpoint - the maps are plotted from their respective vantage points, hence the arcsec scales are different. The grey marked line in panel a shows the solar limb as viewed from Solar Orbiter. In panel (c), the STIX lightcurves are plotted with the five vertical lines corresponding to the five EUV images plotted in (a) and (b). In panel (d) the EUV filament eruption is shown at a later time when the X-ray emission has ceased.}
    \label{fig:euv_images}
\end{figure*}

The source region of the solar eruptive event (i.e. solar flare) was occulted from the three remote sensing viewpoints of STEREO-A, Solar Orbiter, and Earth as depicted in Figure~\ref{fig:positions}. This is highlighted by Figure~\ref{fig:euv_images}, which shows the eruption as observed in the 304~\AA\ channels of Extreme-UltraViolet Imager (EUVI) on-board STEREO-A \citep{euvi} and Solar Orbiter/EUI/FSI. EUI's imaging cadence at this time was 2 minutes. Therefore, each EUI image is paired with the STEREO image closest to it in time, after accounting for the different light travel times from the eruption to the two observatories.  The five frames shows the filament as it comes into the field of view of both instruments, first in STEREO-A/EUVI (row a) and then in EUI/FSI (row b). It should be noted that the scale in arcsec difference is due to the fact that Solar Orbiter and STEREO-A are at different AU distances, at 0.72~AU and 0.96~AU respectively. Panel c presents the STIX light curve, with vertical dashed lines correlating the timing of the frames to specific points on the light curves. A comparison of the FSI frames with the STIX light curves distinctly reveals an increase and subsequent decrease in X-ray emission as the filament eruption becomes visible in EUI/FSI (and hence STIX), matching the CME's outward expansion. Panel d further shows the prominence's continued expansion and acceleration, captured by the large field of view of FSI.

To estimate the 3D location and propagation angle of the eruption, we applied a line-of-sight triangulation technique \citep{ryan_2024} to the isolated feature in the core of the eruption.
This is clearly identifiable in both EUVI and FSI at 21:56~UT (Figure~\ref{fig:euv_images} (a, b), 4th column). This analysis revealed that the eruption originated approximately 59$^{\circ}$ behind the solar limb as viewed from Earth - a finding consistent with \cite{mierla_2022} - and about 45$^{\circ}$ by the limb from the perspective of Solar Orbiter. Assuming the eruption propagated radially, the occultation heights are estimated at around 250~Mm. This implies that any X-ray emissions observed from Solar Orbiter must originate from an altitude exceeding 250~Mm, or at least 0.3~\(R_\odot\), above the solar limb. This supports our assertion that X-ray emission observed by STIX must originate in the filament eruption itself, as flare footpoints and loop arcade would be too low to be visible above the limb.

\begin{figure}
    \centering
    \includegraphics[width=0.40\textwidth]{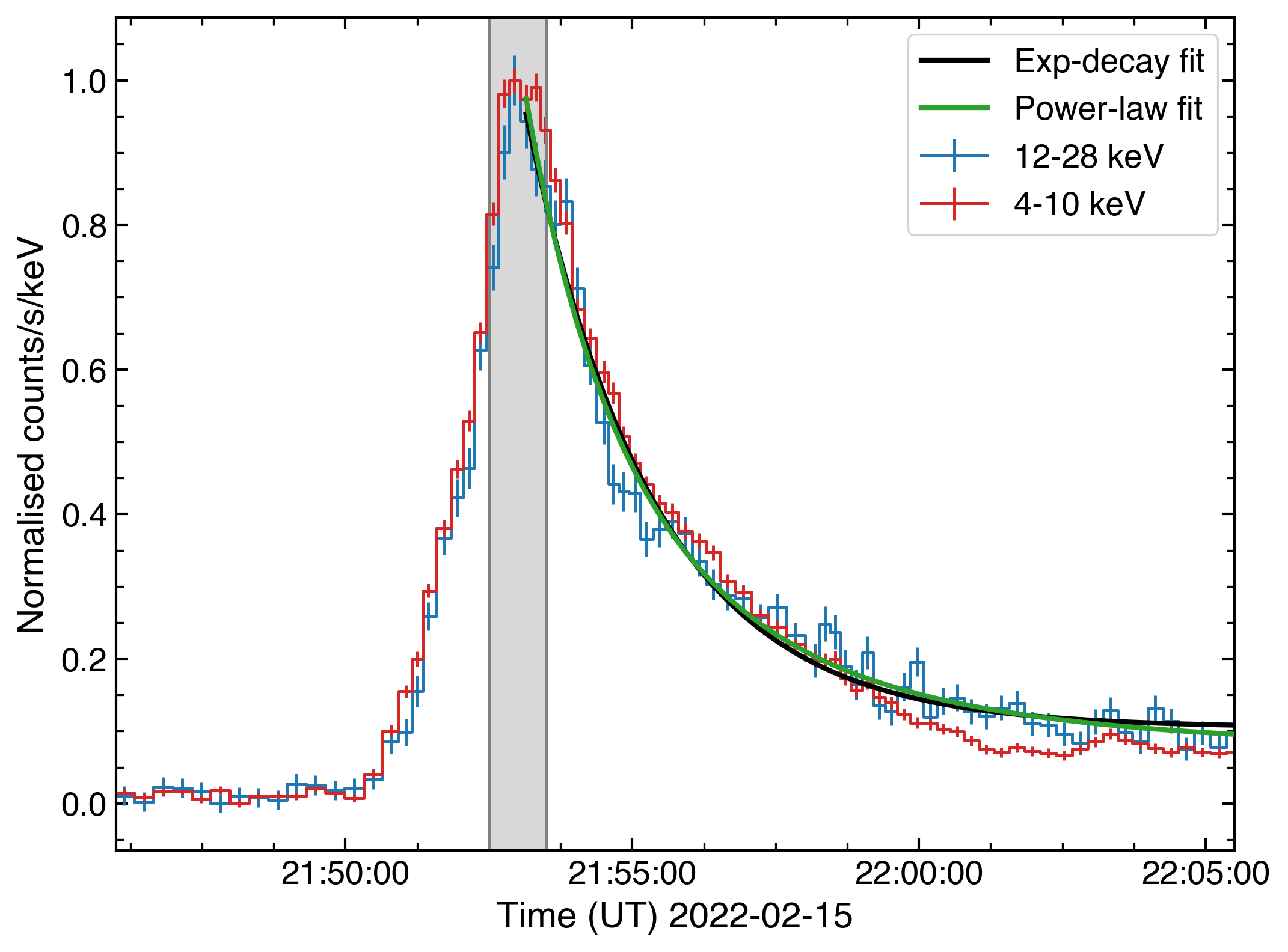}
    \caption{The STIX 4-10 keV and 12-28 keV time-series plotted (normalised). The grey shaded region marks the time interval over which the spectral fitting is performed. The black marked curve is the exponential-decay fit to the decay. }
    \label{fig:exp_decay}
\end{figure}

The time profiles of the STIX X-ray emission show a sharp rise and an exponential decay. The time evolution profile is due to a combination of actual variation of the X-ray emission and an increase of the visible part of the source (see Figure~\ref{fig:euv_images}). The time-profiles across energies are quite consistent which is highlighted further in Figure~\ref{fig:exp_decay}, where the 4-10~keV and 12-28~keV normalised lightcurves are shown. The decay appears similar in both the thermal and non-thermal emissions (this is determined in Section~\ref{section:spectra}), and fitting an exponential decay to the time-series we find a decay time of $ \tau \sim$ 130~s. This is notably very similar to the case reported in \cite{krucker_2007}. Using this value, and Equation 2 for the collisional energy loss from \cite{krucker_review} together with an energy of 20~keV, we can estimate the electron number density, which we find to be on the order of $10^8$~cm$^{-3}$, which is a reasonable value for the high corona (see more in Section~\ref{section:calculations}). However, we should also note that a power-law decay also fits this decay well, with a power law exponent of $\sim$3.5. The time profiles are consistent with a scenario that the X-ray source comes into view of Solar Orbiter/STIX such that the X-ray emission rapidly increases, followed by a decay as the X-ray source expands outwards.

As we have no direct observation of the flare associated with the eruption, we cannot estimate the GOES-class of the flare. However, from the STIX observations, we can estimate what the corresponding GOES-class would be of the X-ray emission associated with the filament eruption itself. This can be done by using correlations between the STIX 4-10 keV and the GOES/XRS 1--8~\AA\ channel of historically observed events \citep{hualin_stix}. Using this, the GOES-class of the peak emission from the filament is between a B6 and C.1 flare. However it should be noted that this correlation relationship between the GOES/XRS X-ray flux and STIX flux is based upon regular flares rather than X-ray emissions associated with a CME so it is not necessarily clear if this correlation should hold for the case of this observation. For example, if the temperature of the source is very high, it would give a relatively small GOES-class due to the differences in the GOES/XRS and STIX temperature responses. If instead we use the temperature and emission measure from the fit of the STIX spectra (see Section \ref{section:spectra}), the estimated GOES-class from these parameters is that of an A4 GOES-class flare. However, we note here that this value is a lower limit of the estimated equivalent GOES-class of the high coronal X-ray source, as it only contains the hottest temperatures that STIX is sensitive too, and it is most likely that there are also contributions from the `cooler' plasma that GOES/XRS would be sensitive too, but we just do not have the observations in those energy ranges. In either case, it is notable that the STIX fluxes of this event is a substantial level of emission given the eruption's separation from the main flaring site behind the limb. Yet had the event not been occulted, this emission would not have been distinguishable amid the much brighter flare emission

\subsection{X-ray spectral analysis}
\label{section:spectra}

\begin{figure}
    \centering
    \includegraphics[width=0.45\textwidth]{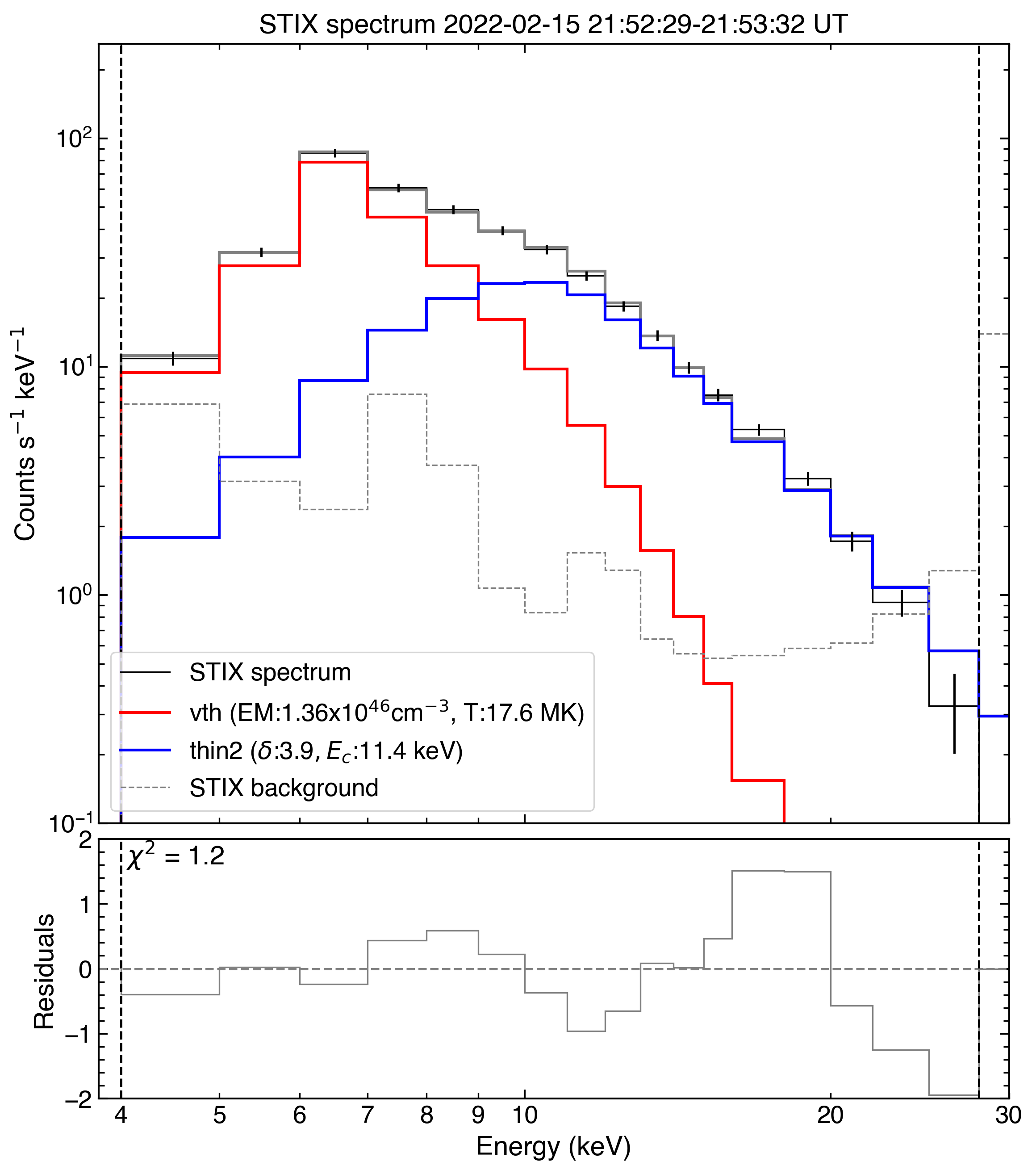}
    \caption{The STIX background subtracted X-ray count flux spectra (black) over the peak of the X-ray emission, and the fitting results. The isothermal fit is plotted in red, and the non-thermal thin-target model is plotted in blue. Below the spectrum is the residuals of the fit to the data. The vertical dashed lines mark the energy range over which the spectral fit was performed.}
    \label{fig:spectra}
\end{figure}

In order to determine the nature of the coronal X-ray emission observed by STIX, we conducted spectral analysis focusing on the peak of the X-ray emission. The time-interval used here is within the grey shaded region in Figure~\ref{fig:exp_decay}, during the peak minute from 21:52:29 to 21:53:32 UT, ensuring a sufficient count rate for analysis. The background-subtracted\footnote{Background file from quiet period on 2022-02-16, UID: 2202160007} spectrum was then fitted using \code{OSPEX} \citep{ospex_2020}. The spectrum is well represented by an isothermal model (\code{f\_vth}) combined with a thin-target non-thermal model (\code{f\_thin2}) fitted over an energy range of 4 to 28~keV. We attempted to fit different models to the data, including a super-hot component, and combinations of thick-target and thermal components, but found the isothermal and non-thermal thin-target combination of models fit the best. The results of the fit are shown in Figure~\ref{fig:spectra}, where the background-subtracted X-ray count spectra is plotted in black, and the thermal and non-thermal models are shown in red and blue, respectively. The vertical dashed lines show the energy range over which the spectrum was fit, and the bottom panel shows the residuals of the fit to the data. Below approximately 10 keV, thermal emission is dominant, with the fit indicating a low energy cutoff at 11.4$\pm$0.4~keV. The spectra also reveal a pronounced non-thermal component, characterised by an electron spectral index ($\delta$) of 3.9$\pm$0.2. The fit of the thermal component suggests a  high temperature of 17$\pm$2~MK and a relatively low emission measure of $1.36\pm0.56 \times 10^{46}$~cm$^{-3}$, consistent with the emitting plasma being a hot, diffuse X-ray emitting source.

\subsection{X-ray imaging}
\label{section:imaging}

STIX is an indirect Fourier imager, and uses a grid-based approach to reconstruct images from spatial modulation patterns of the incident X-ray flux on the detectors \citep{krucker2020, massa_2023}. The instrument is designed such that the grid pitches and orientations provide spatial information on angular scales on the Sun of 7\arcsec to 180\arcsec. To generate images from STIX visibilities, significant flux and signal modulations are essential from each subcollimator. Utilising the forward fit algorithm \citep[FWDFIT;][]{volpara_2022}, we constructed a sequence of thermal X-ray images with 30-second integration in the 4-10 keV range during the peak X-ray emissions. Modulations were primarily detected in the coarsest grids (subcollimators 7-10), indicating an extended source size, which were then employed for image construction. For each time interval, the CLEAN algorithm was also used to construct the X-ray source images, adopting a clean beam width of 60\arcsec, corresponding with the resolution of the finest grids used (i.e. subcollimator 7).

\begin{figure*}
    \centering
    \includegraphics[width=0.9\textwidth]{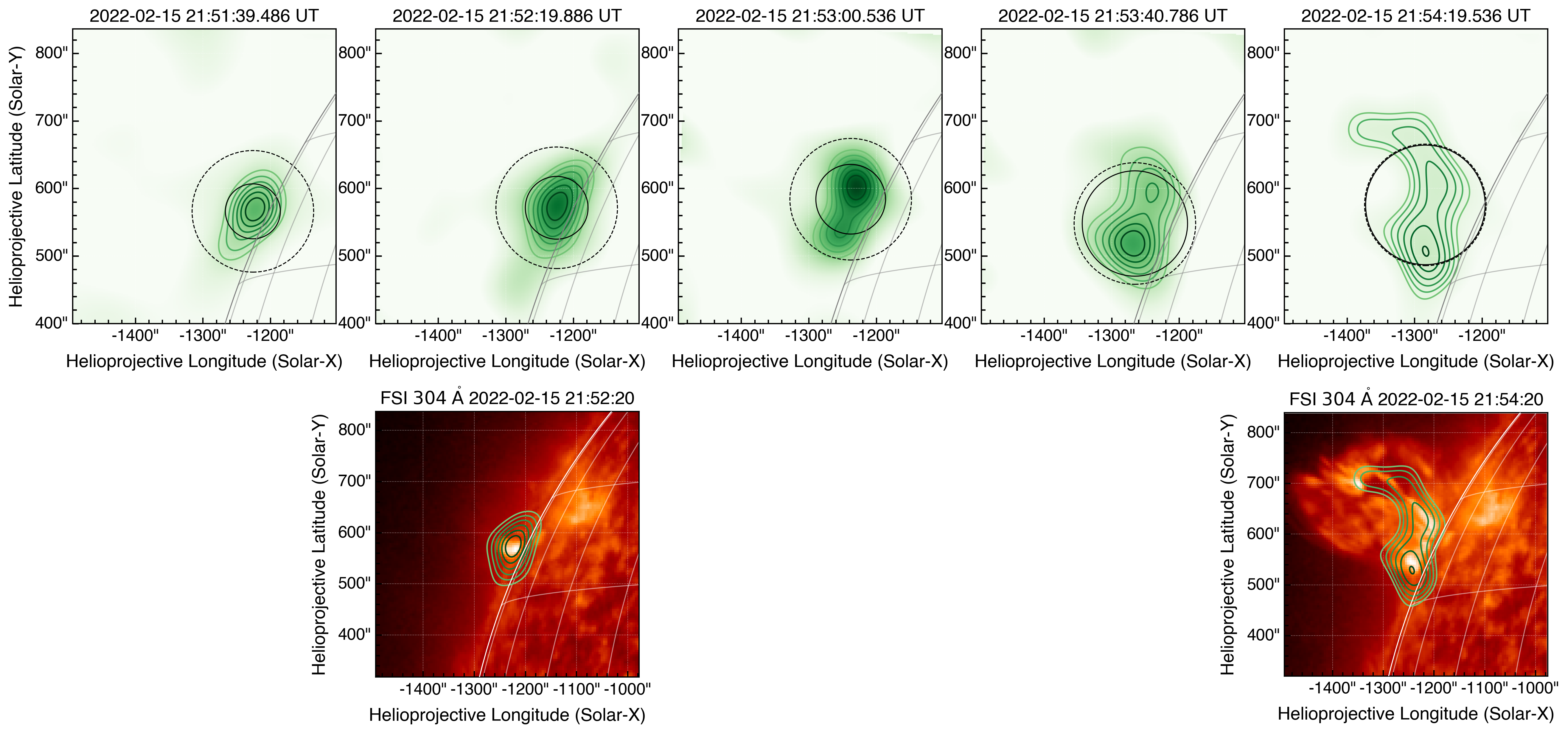}
    \caption{The time-evolution of the X-ray 4-10~keV source during the peak of the X-ray emission. The top panel shows the STIX images reconstructed using the CLEAN algorithm with a 30s time integration. The green colormap, and green contours show the CLEAN intensity maps, scaled to the peak flux, and the contours show the 40-100\% levels. The black circle marks the FWHM of the FWDFIT reconstructed maps over the same time intervals and the dashed black circle marks the 180~\arcsec limit. The corresponding EUI/FSI maps for the two time intervals for which  observations are available are plotted below with the STIX contours overplotted on-top. To overplot these STIX images on the corresponding EUI/FSI maps, the contours are overplotted using the `\code{SphericalScreen}` context manager available through the \code{sunpy} package. For the right-most panel, the STIX contours are shifted by (40~\arcsec, 20~\arcsec) to align better with the EUV emission based on the brightest sources.}
    \label{fig:xray_images}
\end{figure*}

Figure~\ref{fig:xray_images} presents the reconstructed STIX thermal X-ray images across five intervals at the X-ray emission peak in the top panel row. For the two time intervals for which a corresponding EUI/FSI map is available, the map is plotted directly beneath. The CLEAN reconstructed maps are denoted by the green colormap and contours at each time-step, with the forward fit map full-width at half maximum (FWHM) represented by a black circle. The dashed black circle delineates the 180" boundary limit of the coarsest STIX sub-collimator. The observed thermal X-ray source in the 4-10~keV energy range is large and extended, aligning with the filament eruption's entry into Solar Orbiter’s field of view. The FWDFIT FWHM indicates the source's growth and outward movement, reaching the 180" limit of the coarsest sub-collimator, as visible in the final panel. After this time, we cannot reliably construct an image with STIX, as the source is too extended. The CLEAN maps show an evolution of the source that similarly follows the FWDFIT maps, but also shows structure as it evolves which appears to track the legs of the filament eruption. We find similar morphology using the MEM\_GE imaging algorithm. Given the limited temporal resolution of EUI/FSI we can only compare the time the source comes into view and then notable in the last panel where the thermal source mirrors the EUV structure's brightest “arm” feature, suggesting that the bright emission in EUV 304~\AA\ also has a hot thermal component as evidenced by the STIX X-ray source. For the last panel, we shifted the STIX image in this frame by (40\arcsec, 20~\arcsec) to match with the brightest sources of the EUV emission. This is similarly observed in the 174~\AA\ channel of FSI. 

\begin{figure}
    \centering
    \includegraphics[width=0.45\textwidth]{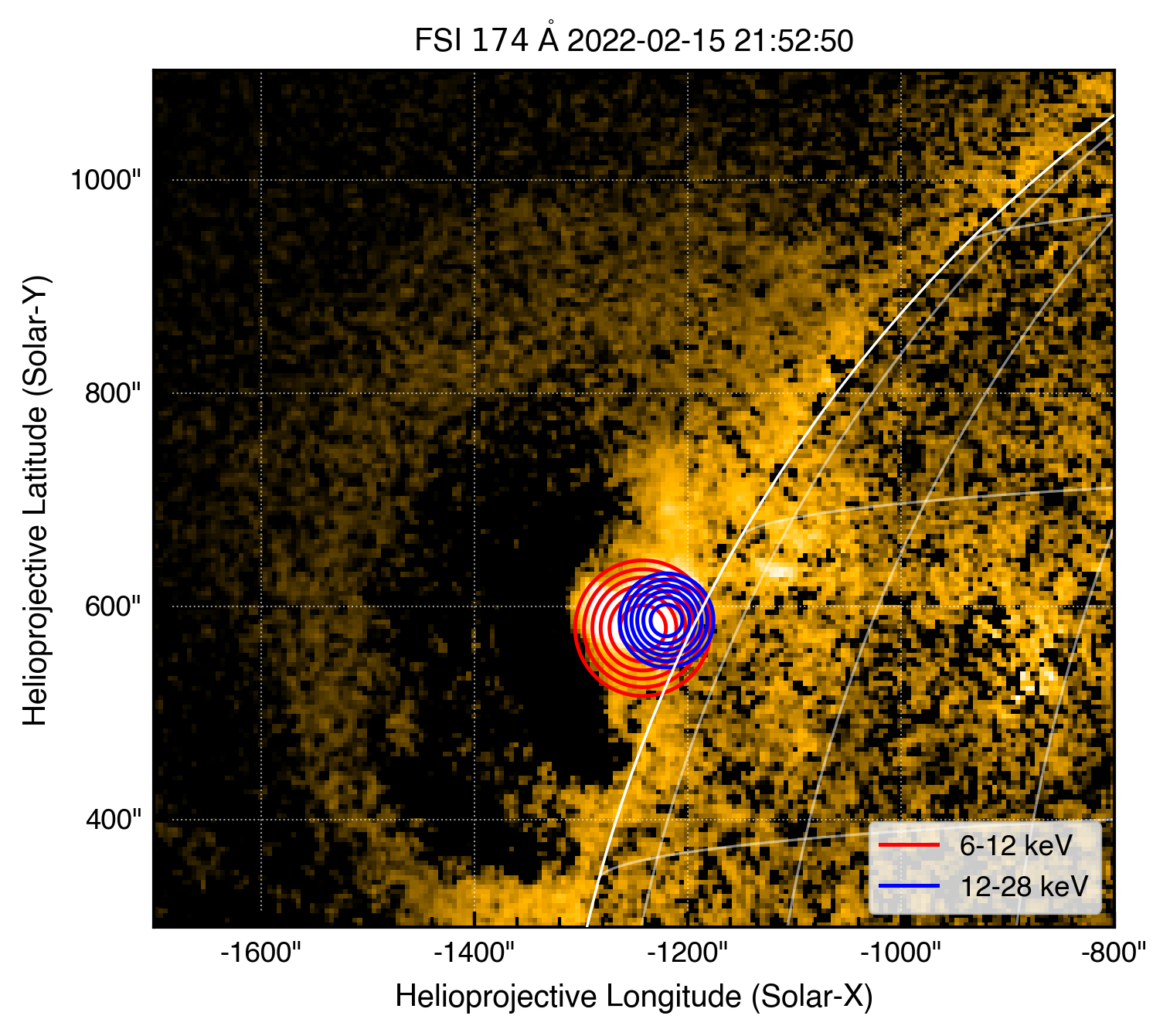}
    \caption{The non-thermal 12-28~keV (blue) and thermal 4-10~keV (red) forward fit X-ray sources over the peak of the X-ray emission overplotted on the running difference EUV 174~\AA\ image from EUI/FSI.}
    \label{fig:xray_images_peak}
\end{figure}

For the non-thermal emissions, we choose an energy range from 12-28~keV to construct an image. This was chosen as the fitted count spectrum shows that the energy range from 12-28 keV is largely non-thermal. Given that the count statistics are lower relative to the thermal X-ray source, we use a longer integration time of 1 minute over the peak of the lightcurve to construct an image. For reference, we also construct a thermal source image in the 4-10~keV over the same time range interval. We applied the forward fit algorithm (FWDFIT) to obtain estimates of the positions and sizes of both thermal and non-thermal sources during the peak X-ray emission. Figure~\ref{fig:xray_images_peak} displays these, with the 12-28~keV non-thermal source outlined in blue contours and the 4-10~keV thermal source in red. These contours are overplotted on a running difference image from EUI/FSI at 174~\AA\, which distinctly shows the CME shock front or the leading edge of the CME, prominently visible well above the filament eruption and the X-ray sources. According to the forward fit results, the thermal source exhibits a FWHM of 110$\pm$16\arcsec while the non-thermal source measures 76$\pm$17\arcsec. Although their locations slightly diverge, pinpointing their precise positions relative to the filament eruption and its orientation presents a challenge. However, it seems that both sources are situated within the "legs" of the filament as it enters Solar Orbiter's field of view.

\subsection{Energetics of the X-ray sources}
\label{section:calculations}
Combining the parameters derived from the X-ray spectral fitting, and X-ray imaging we can estimate the thermal source properties. We determine the volume of the X-ray emitting thermal plasma through the forward fitting of the circular source over the peak of the emission from Figure~\ref{fig:euv_images}. Here, after using the FWDFIT source to estimate the volume, we also tested the method using CLEAN maps by estimating the area within the 30\% contour levels and found a similar volume estimate. We can then estimate the ambient density, $n_e$ of the thermal source by combining with the emission measure (EM) derived from the spectral fit;
\begin{equation}
n_e = \sqrt{\frac{EM}{V}} \qquad \unit{[cm^{-3}]}
\end{equation}
If we assume the simple approach that the thermal source is spherically symmetric with a filling factor of one, (i.e. $V=fV$, where $f<1$), the derived density of the hot (17~MK) plasma source is $\sim 3.7 \times 10^8$~cm$^{-3}$; a reasonable value for CME high in the corona. This source hence contains around $3.8 \times 10^{37}$~electrons, and has an estimated thermal energy content of $\sim 2.7 \times 10^{29}$~erg. We want to note that these values are estimates, rather than definitive values. For example the largest uncertainty in these calculations is the volume estimate. If we use the ranges of volumes given from the uncertainties in the source size, we get ranges of the $n_e$ to be between $(3-4.7) \times 10^8$~cm$^{-3}$, the number of electrons in the thermal source to be between $(3-4.6) \times 10^{37}$, and the thermal energy content between $2.1-3.2$~ergs.

\begin{figure}
    \centering
    \includegraphics[width=0.40\textwidth]{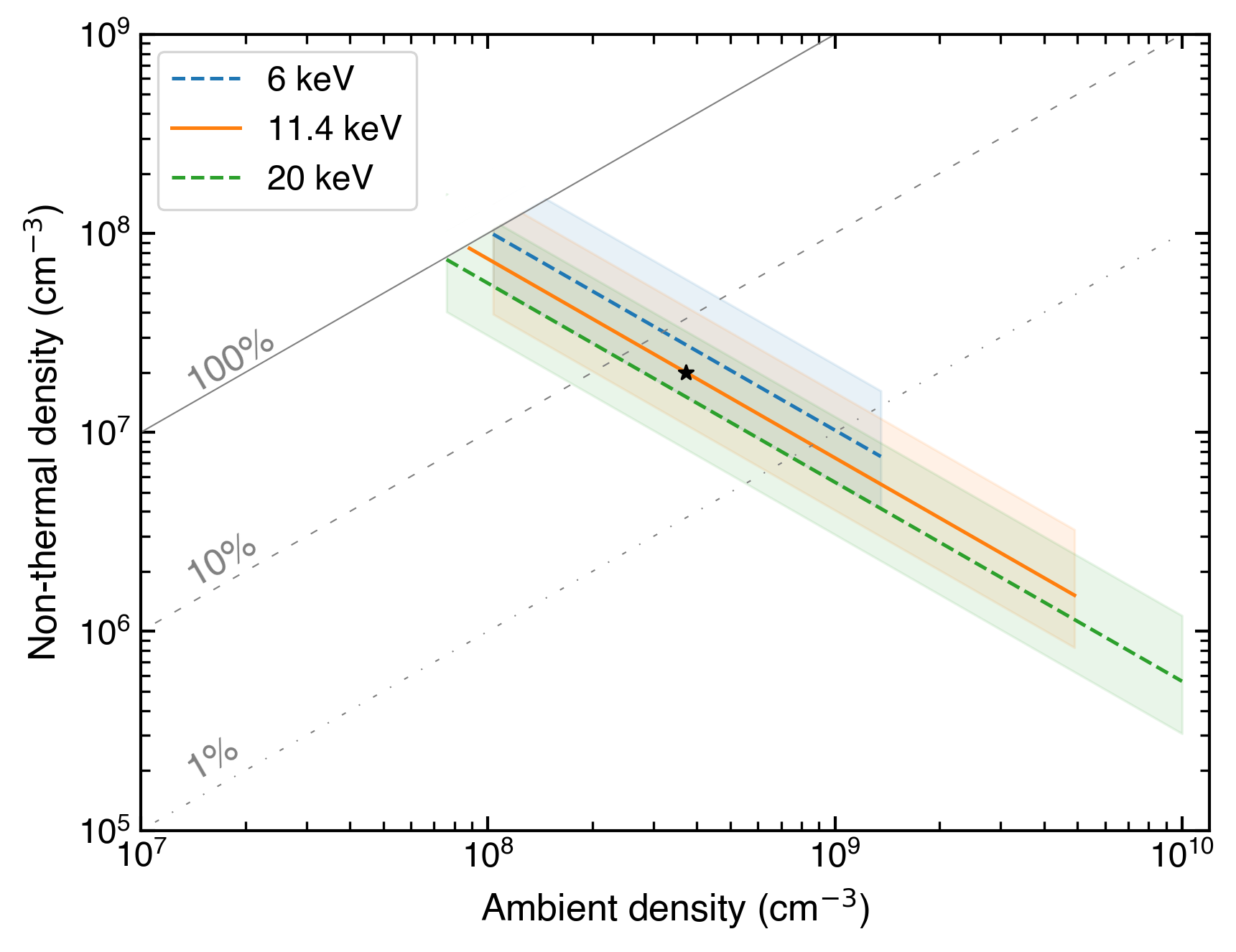}
    \caption{A plot of the instantaneous non-thermal electron density as a function of ambient density for three different low-energy cut-offs. The instantaneous non-thermal electron density is only calculated for which the thin-target assumption holds. The shaded regions mark the ranges of the volume calculation from the source size. The grey lines at 100\%, 10\% and 1\% mark where the ambient density equals the respective percentages of the non-thermal density (i.e. 100\% is the extreme cases when the ambient density equals the non-thermal electron density.) The black star marks the value of the instantaneous non-thermal electron density if we assume the ambient density is that of the thermal source.}
    \label{fig:nonthermal_den}
\end{figure}

For the non-thermal emission, we can estimate the instantaneous electron density and number of non-thermal electrons. For a thin-target model, this can be estimated by using Equation 6 from \cite{kontar2023}, \citep[see also Appendix B in][]{musset_2018}, given by; 
\begin{equation}
n_{nth} = \frac{\langle n V F_0 \rangle}{n V_{nth}}  E_0^{-0.5} \frac{\delta-1}{\delta-0.5} \sqrt{\frac{m_e}{2}}   \qquad \unit{[cm^{-3}]}
\end{equation}
 where $n$ is the ambient density, $\langle n V F_0 \rangle$ and $\delta$ are the normalisation factor and the electron spectral index from the thin-target model fit, respectively, $E_0$ is the low energy cut-off, $V_{nth}$ is the non-thermal source volume, and $m_e$ is the electron mass in keV/$c^2$. From the spectral fit from Section~\ref{section:spectra} ($\langle n V F_0 \rangle = 1.86 \times 10^{54}$~electrons~s$^{-1}$~cm$^{-2}$, $\delta_{thin}$ = 3.9, and $E_c$ = 11.4~keV), and estimating the volume from Section~\ref{section:imaging} (V$_{nth} \sim 3 \times 10^{28}$ cm$^3$, assuming spherically symmetric source with filling factor of unity). As the ambient density is not well constrained, and similarly the low-energy cut-off is not well constrained, we can plot the instantaneous non-thermal electron density for a range of ambient densities
and low-energy cutoffs, in a similar manor to that done in \cite{lastufka2019}. For a range of ambient densities, we only calculate the instantaneous non-thermal electron density for values for which the thin-target assumption holds (i.e. the ambient density does not act as a thick-target). To determine the high densities for which the assumption does not hold, we calculate the stopping column densities for different cut-off energies using the estimated source size from Section~\ref{section:imaging}, and Equation~1 from \cite{krucker_review}. In Figure~\ref{fig:nonthermal_den}, this is shown for three ranges of the $E_c$, namely 6~keV, 11.4~keV and 20~keV. The shaded regions note the ranges of values for the non-thermal volume estimate, V$_{nt}$. The grey solid line shows the extreme case when the ambient density equals the non-thermal density, and the dashed grey lines show the cases when the non-thermal density is 10\% and 1\% of ambient density.  If we take  ambient density, $n$, to be that of the thermal source ($\sim 3.7 \times 10^{8}$~cm$^{-3}$), we get an estimated $n_{nth} \sim 2 \times 10^7$~cm$^{-3}$ (marked by the black star in Fig~\ref{fig:nonthermal_den}). With these assumptions, we find that the ratio of the non-thermal to thermal electron density is $\simeq$ 0.05. Under these assumptions, that the ambient density is the thermal source, we find a total number of non-thermal electrons of $\sim 7 \times 10^{35}$, which is approximately 2 \% of the total population of electrons, and a total energy content of the non-thermal source to be $\sim 2 \times 10^{28}$~erg. This is approximately 7~\% of the estimated thermal energy. Similar to the thermal source estimates, if we take the ranges of the volume estimates for the non-thermal source, we find that the total number of non-thermal electrons range from ($5.3-8.3) \times 10^{35}$, and the total energy range from $(1.5-2.3) \times 10^{28}$ ergs.

\section{Discussion \& conclusion}
The large filament eruption and CME that occurred on February 15 2022 provided an excellent opportunity to study an occulted solar eruptive event from the perspective of Solar Orbiter. By combining STIX and multi-viewpoint EUV observations from both EUI/FSI and STEREO-A/EUVI, this study reveals the existence of a hot (>17~MK) X-ray component and the presence of flare-accelerated non-thermal electrons within the coronal source associated with a filament eruption located at least 250~Mm radially above the main flare site. The X-ray time-profile is relatively simple; it peaks as the filament comes into view of Solar Orbiter, as identified in EUV from EUI/FSI, and shows an exponential decay on the timescale of $\sim$130~s. The X-ray source was imaged over the peak of the emission, and shows a source that spatially aligns with the EUV structure as seen in the 304~\AA\ and 174~\AA\ channels of EUI/FSI, and grows in size until it reaches the detection limit of STIX’s largest grids of 180\arcsec, for which after an X-ray image cannot be reconstructed as its too extended to show any modulation even in the coarsest grids of STIX. The presence of the hard X-ray non-thermal emissions above 11~keV, with an electron spectral index of 3.9$\pm$0.2~keV, suggest that there is a population of accelerated electrons that are confined to magnetic field line structures associated with the CME and filament eruption, likely situated in the legs of the filament eruption that is observed in EUV, behind the CME shock. 

The question then is, how the X-ray emitting plasma of the filament reached temperatures as high as 17~MK, and how a population of accelerated electrons became embedded within this structure at these altitudes? A likely scenario, based on a 3D eruptive flare model \citep[e.g.][]{janvier_2015}, is that flare-accelerated electrons escaping upwards from the acceleration region near the current sheet are injected into the complex magnetic configuration of the filament. Within this structure, they are then trapped and carried outward with the erupting filament. Recent observations have documented similar processes at lower altitudes - microwave emission from accelerated electrons at the conjugate legs of a filament eruption \citep{chen2020} and hard X-ray detection at the filament anchor points in the chromosphere \citep{stiefel_2023}. This study extends those findings to their coronal counterparts, suggesting that these electrons can remain trapped within the filament structure for an extended duration, observable at these higher altitudes. It should also be noted that the detection of Type-III radio bursts indicates some flare-accelerated electrons have also escaped upwards onto open magnetic field lines, as well as those that access the confined magnetic structure.

The hot thermal X-ray sources match with the structure observed in the EUV channels of EUI/FSI, and are similarly seen in the other hotter coronal channels from EUVI \citep[see Figure 5 in][]{mierla_2022}. Although the FSI 304~\AA\ passband is typically dominated by He II 304~\AA, it also contains hotter lines indicative of temperatures around 2 MK, and the FSI 174~\AA\ channel has a peak temperature response at approximately 1 MK. These observations suggest that the filament exhibits a multi-thermal distribution and is significantly heated. The heating of the flux rope to temperatures up to tens of MK could be attributed to flare-accelerated electrons heating the CME filament, a mechanism that has been proposed in similar studies \citep[see][]{glesener2013}. We strongly encourage modellers 
to further investigate the injection of accelerated electrons into different plasma conditions \citep{galloway} and in particular the injection in CMEs structures. This will allow us to understand the production of bremsstrahlung emission from these scenarios.

The observed time-profiles of both thermal and non-thermal X-ray emissions, characterised by a sharp rise followed by an exponential decay, also offers insights into our understanding of the nature of the sources. Typically, in emission associated with a solar flare, one would expect the non-thermal emissions to peak sharply and decay rapidly, preceding the thermal emissions which would peak later and decay more gradually. However, in this case, both the thermal and non-thermal emissions exhibit similarly structured profiles and decay rates. The non-thermal decay could be explained by collisional losses; however, different energy levels would likely exhibit varying decay times under such a scenario. Another plausible explanation is that as the emitting source moves outward, the ambient density decreases, leading to a flux decay. Since both thin-target and thermal bremsstrahlung emissions are dependent on ambient density, an isotropic expansion of the emitting volume would naturally result in a decrease in emission over time, proportional to the cube of the expansion time, aligning with our observations. In this way, the observations of the initial sharp rise in X-ray emissions is due to the emitting volume moving into Solar Orbiter’s field of view from behind the limb. Once fully observable by STIX, the source continues to expand outward, leading to the observed decay in emission. This can also explain the absence of detectable X-ray emissions when the event becomes visible from Earth; by this stage, the expanded volume would be too diffuse such that emission measure below the detection threshold of GOES/XRS. In reality, both  collisional losses and the expansion are at play, but overall the observations show a consistent picture of hot plasma and trapped non-thermal electrons within a filament structure. The non-thermal population observed is a lower limit of the total electrons in the high-altitude source, and there may have been more at earlier altitudes that could have contributed to the heating of the structure. This study shows that while the emission is weak, they are still energetically significant, and these studies are important for understanding how the energy budget is utilised in solar eruptive events. 

Future work will leverage STIX observations in conjunction with Earth-based observatories, including the newly operational Hard X-ray Imager \citep[HXI;][]{zhang_hxi} on-board ASO-S \citep{asos}, the Gamma-ray Burst Monitor (GBM) \citep{meegan_gbm} and Aditya-HEL1OS, to systematically search for solar eruptive events from multiple vantage points to investigate. This approach will enable us to identify configurations where the main flare is occulted from one viewpoint but visible from another. Such multi-viewpoint observations will allow simultaneous X-ray analyses of both the flare and the coronal CME component, providing a comprehensive view of the high-energy aspects of solar eruptive events. In addition, we will attempt to search for events that occur during the perihelion of PSP, such that PSP measures the in-situ aspect energetic electrons of the CME very close to the Sun that could be compared with the remote sensing X-ray observations. 

While this study demonstrates the capability of STIX to measure coronal X-ray emissions associated with a CME, the identification of the source was facilitated by the fact that the main flare was occulted behind the limb. The use of indirect Fourier imagers like STIX are inherently limited by the dynamic range of the observations, requiring reliance on occulted events - where the bright flare source is obscured - to detect coronal X-rays. Looking to the future, the adoption of a direct X-ray imager, such as FOXSI \citep{christe, foxsi}, which has been proposed for as part of missions concepts of FIERCE \citep{fierce} and SPARK \citep{spark}, would revolutionise our observational approach. A direct imager would allow for simultaneous imaging of both the flare and the coronal emission, substantially improving our understanding of the complex dynamics of solar eruptive events and allow us to identify the locations of particle acceleration and heating and whether electron acceleration occurs in the current sheet below the CME or within the CME itself. 
 
\begin{acknowledgements}
Solar Orbiter is a space mission of international collaboration between ESA and NASA, operated by ESA. The STIX instrument
is an international collaboration between Switzerland, Poland, France, Czech Republic, Germany, Austria, Ireland, and Italy. L.A.H is supported by an ESA Research Fellowship. S.K. and H.C. are supported by the Swiss National Science Foundation Grant 200021L\_189180 for STIX. The authors thanks the anonymous referee for their helpful comments which improved this manuscript.
\end{acknowledgements}

  \bibliographystyle{aa.bst} % style aa.bst
  \bibliography{aanda.bib} % your references Yourfile.bib

\begin{appendix}
\section{Software used}
This paper made use of several open source packages including astropy \citep{astropy:2022}, sunpy \citep{sunpy_community2020, Mumford2020}, matplotlib \citep{Hunter:2007}, numpy \citep{harris2020array}, scipy \citep{2020SciPy-NMeth}, pandas \citep{reback2020pandas}, sunpy-soar. The IDL software used in this work used the SSW packages, and version v0.5.3 of the STIX software. See also https://github.com/i4Ds/STIX-GSW.

% \section{Calculations used}
% The collisional energy loss time of an electron of energy  E in keV is given by 
% \begin{equation}
%     \tau(E) = A \frac{E^{3/2}}{n_e}
% \end{equation}
% where $n_e$ is the electron number density and A = $2\times 10^8$ s keV $^{1.5}$ cm$^{-3}$.

% The STIX data used in this paper consists of both:

% \begin{itemize}
%     \item 0.5s pixel data with UID: 2202159188
%     \item 10s pixel data with UID: 2202154670
%     \item 0.5s spectrogram data with UID: 2202150035
%     \item background data file with UID: 2202160007
    
% \end{itemize}

% \begin{figure*}
%     \centering
%     \includegraphics[width=0.9\textwidth]{figures/clean_maps_174_410_7.png}
%     \caption{The same as Figure~\ref{fig:xray_images}, except with the 174~\AA\ channel of EUI/FSI.}
%     \label{fig:xray_images2}
% \end{figure*}

\end{appendix}


% \begin{thebibliography}{}

%   \bibitem[Baker(1966)]{baker} Baker, N. 1966,
%       in Stellar Evolution,
%       ed.\ R. F. Stein,\& A. G. W. Cameron
%       (Plenum, New York) 333

% \end{thebibliography}

\end{document}